# A SDN-based Flexible System for On-the-Fly Monitoring and Treatment of Security Events

Joberto S. B. Martins
Salvador University, UNIFACS, Brazil
joberto.martins@unifacs.br

Maxli B. Campos
Salvador University, UNIFACS, Brazil
mcampos.2004@gmail.com

*Abstract*— **The Software Defined Networking (SDN) paradigm decouples control and data planes, offering high programmability and a global view of the network. However, it is a challenge not only provide security in these next generation networks as well as allow that network attacks could be subjected to an incident and forensic treatment procedure. This paper proposes the implementation of flexible mechanisms of monitoring and treatment of security events categorized per type of attack and associated with** *whitelist* **and** *blacklist* **resources by means of the SDN controller programmability. The resources to perform intrusion and attack analysis are validated by means of a real SDN/OpenFlow testbed.**

## I. INTRODUCTION

Current computer networks do not meet some newer requirements and present new demands that need to be rethink. According to [1] the absence of flexibility in controlling the internal operation of the equipment as well as the high cost of existing infrastructure are barriers to the evolution of architecture and the necessary innovation in the provision of new services and network applications. One of the initiatives in this direction is the Software Defined Network (SDN) paradigm that is in most cases based on the OpenFlow protocol [2]. SDN typically employ a controller to install, on demand, packet forwarding rules per flow in the network nodes [3]. Despite the advantages brought by the SDN networks, some of the vulnerabilities of traditional networks persist in SDN-based infrastructures, such as the typically centralized nature of the control plane, added to the fact that many of the tools and techniques used for security information as anti-virus and firewall are not sufficient to ensure safety [4]. The intrusion detection technologies, adequately integrated with the SDN environment, can provide an additional security element [5].

The use of OpenFlow in network security area considering analysis and intrusion prevention has been discussed and presented good results in previous research. For example, Lopes et al. [6] proposes BroFlow, an elastic and distributed IDPS (Intrusion Detection Prevention System) for defense against DoS attacks in virtualized SDN, that is based on the OpenFlow API and network traffic analyzer Bro. Jankowski et al. [7] presented a solution where the outcome was based on the assumption that it is possible to classify whether network traffic flows represent normal operation or attack and the flows classification is based on features obtained through the functionality available in the SDN technology. Xing et al. [8] presented the SnortFlow, a proposal for IPS in a cloud environment, with XEN using OpenFlow switches to assist in capturing traffic. In this study, authors developed a prototype in which the SnortFlow agent was installed in the areas Sun 0 and Dom U.

Many of these works cover general features of SDN networks protection without making a demonstration with real network equipment, generating results using either simulation tools or prototypes. Thus, the purpose of this work is, firstly, to extend the related works through the implementation of an additional flexible mechanisms for monitoring and treatment of security events categorized per type of attack and associated with *whitelist* and *blacklist* resources. Secondly, the proposed solution will provide mechanisms to perform intrusion and attacks analysis with validation by means of a real SDN-OpenFlow testbed. Regarding the methodology used, it is characterized as an applied case study that explores the programmability of the SDN controllers in a real testbed. The focus is on validating the proposal itself and its implemented mechanisms for detection and reaction to future use at the Salvador University (UNIFACS) FIBRE network [9].

## II. THE PROPOSED SYSTEM

SDN enables more flexible and predictable network control and makes it easier to extend the network with new functionality through the programmability of the controller. In this context, this work employs Ryu as the application coding platform for the control plane. Ryu is one of the controllers on the market developed by the Center for Innovation in Software (NTT Japan) under Apache 2.0 license [10]. Ryu project is a framework based on software programming components that uses Python as the programming language and allows the development of new applications using multiple SDN protocols, including OpenFlow 1.0, 1.2, 1.3 and 1.4.

As shown in Figure 1, two applications were implemented on top of the Ryu controller. The first is Snort_Switch (*switch_snort.py*) which is the main code supporting a L2 Switch and responsible for enabling the OpenFlow Switch to redirect all traffic in promiscuous mode from one of its ports, set up for the Snort tool itself. The second application, called "mitigation" (*mitigation.py*), aims to address the *Packet_event* messages sent from Snort tool to Ryu controller when it is reported some event associated with an attack signature in the OpenFlow network. These two applications are intended to provide at the control plane the resources needed to implement the mechanisms of detection and response. The *packet_in_handler* is the main method in the *switch.py* file, where are processed all OpenFlow messages. In this method it is expected to allow the creation of flow entries in switches only if the MAC Source of a *Packet_in* message was registered previously in the *Whitelist* and not be in the *Blacklist*. In this method is also developed the treatment for *Packet_event* messages sent by Snort.

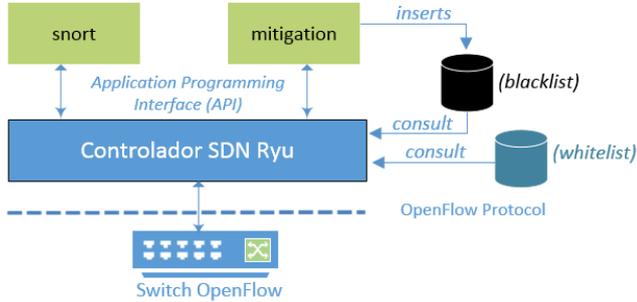

FIGURE 1: RYU-BASED SYSTEM CODE BUILDING BLOCKS

The proposed concept is based on the SDN flow classification mechanism by type of traffic as shown in Table 1. An important aspect of this approach is the test environment for the generation of SDN network traffic, allowing verifying the effectiveness of the treatment mechanisms by type of attack.

TABLE 1: TRAFFIC CLASSES EXISTING IN THE REAL TESTBED [6]

| Traffic Class | Activities | Example of activities | Tools for traffic generation |
|---|---|---|---|
| Class 1 | Normal | Communication between clients and server | FTP, SSH, SMB, Apache, WEB, Postgres, Telnet, Ping ICMP |
| Class 2 | Probe | Port probe, vulnerability and version scan | Metasploit, nmap |
| Class 3 | DoS | Denial of service | Metasploit, hping3, nping |

Table 2 shows the possible actions that have been programmed into the Ryu controller after identifying a suspicious network traffic. Three methods are essential to carry out the treatment of security events set out in Table 2 and allow for each type of class of attack a specific action or set of actions.

TABLE 2: FLEXIBLE TREATMENT OF SECURITY EVENTS

| Traffic Class | Activities | Action | Type of treatment |
|---|---|---|---|
| Class 1 | Normal | Only detection | 1. only storage in database |
| Class 2 | Probe | Detection and Reaction | 1. store in database<br>2. REWRITE flow entries |
| Class 3 | DoS | Detection and Reaction | 1. store in database<br>2. DROP flow entries<br>3. MAC Source in *Blacklist* |

For the types of Class 3 attack the treatment it is given by *remove_table_flows* method, programmed in *mitigation.py* file to send a command to the Switch to drop all flow entries that have the MAC address of the attack source and inserting this MAC in the *blacklist*, blocking the traffic and preventing new connections of the suspect device.

The type of attack Class 2 is considered by *modify_table_flow* method, defined in *mitigacao.py* file, that was programmed to send a command to Switch to rewrite all flow entries that have the MAC address of the attack source and redirect the traffic to port 4 of the Switch connect to a *honeypot* to deceive the attacker and understand (analyze) the used attack signature. In relation to the treatment of attack traffic Class 1, they are processed by Snort itself. In this case, they are detected by means of Snort rules and only performs the action of storing all traffic in the database such that they can be analyzed in the future using the Basic Analysis and Security Engine (BASE) tool.

### III. REAL SDN NETWORK TESTBED

The SDN network in Figure 2 shows the equipment and framework components necessary to implement and validate the flexible Intrusion Detection and Treatment System (IDTS) proposed.

#### A. Description of the SDN network testbed

The experimentation network shown in Figure 2 uses a TP-Link (**1**) WR1043ND hardware, which is a wireless router approved by OpenWRT.org and [11] to operate as a low cost OpenFlow switch. To support the OpenFlow 1.3 protocol, necessary in the proposal, the equipment had its firmware changed and received the OpenWrt operating system version compiled from version "trunk" of OpenWRT (Barrier Breaker) [12].

The experimentation setup has three different networks: the SDN network |IP 172.16.10.0/24| which basically connects the hosts in the OpenFlow network. For the test it was used two real machines as network clients, being one with the operating system Linux Kali |IP 172.16.10.251| (**2**) connected to port 4 of the Switch OpenFlow to simulate some attacks. The other machine is the one under attack (**3**) with Ubuntu Server 16.04. It was also employed one honeypot machine (**4**) emulating machines on the network by means of the *honeyd* tool, a low interaction honeypot. Machine server (**5**) runs the OpenFlow network Ryu controller |IP 192.1681.130/24|. The experimental setup includes protection mechanisms associated with *whitelist* (**6**) and *blacklist* (**7**) features, to demonstrate the effectiveness of the solution. The Ryu controller supports a Switch L2 with integration features with Snort tool using the controller programming flexibility. The Snort machine (**8**) has *eth0* interface configured in *promiscous mode* (**9**) that receives all traffic on port 3 of the OpenFlow Switch in mirror mode (**10**), and *eth1* configured on the |IP 192.168.1.120| to send packets of the type *Packet_event* (**11**) to Ryu controller when a malicious traffic is identified in the network.

The detection and reaction mechanisms are implemented by coding at the file *mitigacao.py* in the controller which handles events *Packet_event* generated by Snort and that are sent to the Ryu controller through an application developed in python called *pigrelay* (**12**) available in [13]. The *pigrelay* tool is running on the Snort machine and when executed sends to the controller in the |IP 192.168.1.130| and port |51234| - security events being generated by Snort tool and stored in */var/log/snort/snort_alert*. To enable the storage of data

generated by Snort tool it was used a MySQL database manager. For security reasons, all access to the network controller is made by a single management console **(13)**.

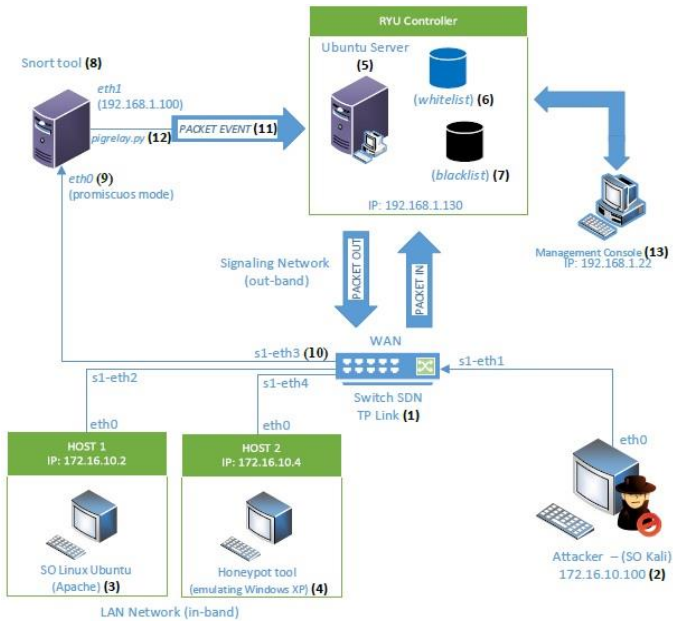

FIGURE 2: REAL SDN-OPENFLOW TESTBED

IV. VALIDATION OF THE PROTECTION MECHANISM

This section presents a brief description of the strategies adopted and implemented in the real SDN Network testbed and the validation of the protection mechanisms. The results obtained in the tests will be presented as command outputs of the Ryu controller that was running in verbose mode. The BASE tool is used as analysis tool for the alerts generated by Snort and available on the real SDN network testbed. For all attack simulation detected by Snort, an alert by means *Packet_event* message is sent to the network controller. The controller by means of application *mitigation.py* extracts alert information seeking the flow that matched with the alert the alert received.

A. Whitelist and blacklist

The creation of a flow entry in the switch will only succeed if the source MAC address is previously registered. As shown in Figure 3, for example, a device with a MAC address |68: 5b: 35: b4: fc: bf| when requesting the creation of an entry flow by means a *Packet_in* message had its request denied and was told to contact the administrator network. The authorized MAC address (Figure 3) are registered previously in *whitelist.txt* file in the directory of the Ryu, read and cached in a list created when executing the *switch_snort.py* file.

FIGURE 3: BLOCKING OF UNAUTHORIZED MAC ADDRESS

The *blacklist* is implemented by the *blacklist.txt* file available in Ryu code folder. It records all MAC addresses that have had some kind of activity considered suspect in the OpenFlow network previously. The insertion of new addresses in *blacklist.txt* file is made in the treatment of Class 3 security events and allows to block new access for the node that originated an attack. At runtime, when a device requests the creation of a flow entry to the controller, a method in *mitigation.py* file is programmed to read the *blacklist.txt* file and identify if the MAC address of the device is on the list. If a device is requesting the creation of a flow entry through a *Packet_in* message, for example, the message is not handled by the controller and the MAC address is considered suspect, as shown in Figure 4.

FIGURE 4: BLOCKING A SUSPECT MAC ADDRESS

B. Class 1 attack

For traffic generation, it was used a *ICMP ping* to test normal traffic on the network from the attacker's machine:

#ping 172.16.10.2

As show in Figure 5, the normal traffic is subject to detecting and recording by means of creating a Snort rule classified as type Class 1. The treatment programmed in the controller for any traffic type Class 1 consists only in detect this traffic and record it in database for future analysis using the BASE tool.

FIGURE 5: ICMP DETECTED (NORMAL TRAFFIC)

C. Class 2 attack

For generation of traffic was used the *nmap* tool from the attacker's machine to test probe traffic on the network:

# nmap -sX -O 172.16.10.2

The tests conducted here consist in identify the type of attacks classified in the proposal as Class 2 and that were properly configured in Snort rules in *rules.local* file are being detected and the reaction mechanism consists of a call to *modify_tables_flow.py* method. This method executes the action to submit a *REWRITE* message to the Openflow switch with the objective to rewrite all flow entries that have the MAC address that originated the attack to the port switch 4 connected to a honeypot. When sending the message to the Switch, the method knows the MAC of the device that originated the attack by *Packet_event* message from Snort, which contains all the attack information, which are loaded in *modify_tables_flow.py* method as parameters (Figure 6).

In this way, it is possible to redirect all traffic to the port 4 of the switch redirecting all traffic and effectively implementing a trap for the attacker with the purpose of gathering information about the attack in progress.

FIGURE 6: CLASS 2 DETECTED (PROBE TRAFFIC)

*D. Class 3 attacks*

The *hping3* tool is an injector of packets used to perform stress tests on the network, allowing simulate DoS (Denial of Service) and DDoS (Distributed DOS) attacks. For simulation purposes, we used the following command in the attacker's machine:

# hping3 -S -p 80 --flood --rand-source 172.16.10.2

The test conducted here consists in identify the type of attacks classified in the proposal as Class 3. When sending to the controller (Figure 7) the *Packet_event* message identifying the probe test as Class 3, the controller through *mitigacao.py* code makes a call to *drop_tables_flow.py* method. This method performs the action of sending a *DROP message* to the OpenFlow switch with the purpose to remove all flow entries that have the MAC address that originated the attack.

FIGURE 7: DOS ATTACK DETECTED (DOS TRAFFIC)

When sending the message to the Switch, the method knows the MAC of the device that originated the attack by *Packet_event* message from Snort, which contains all the attack information, which are loaded in *drop_tables_flow.py* method as parameters. As a result, the method removes all entries flow, interrupting the attack. However, only this action is not effective, because new requests for the device to create new flow entries would be made and it would keep the connection to the network. To avoid this, our approach proposes, as a second reaction mechanism for the Class 3 type attack, to associate the DROP action with the insertion of the MAC address that originated the attack on the *blacklist*. In this way, after any kind of malicious action is detected by a particular network node with this node being blacklisted, it becomes virtually impossible a NEW access, since the network policy follows the default "deny-by-default" premise. The estimated time between the identification of the type of attack, sending of *Packet_event* message to the controller and treatment given by *drop_tables_flow.py* method is taking an average of 2 to 3 seconds (Figure 8).

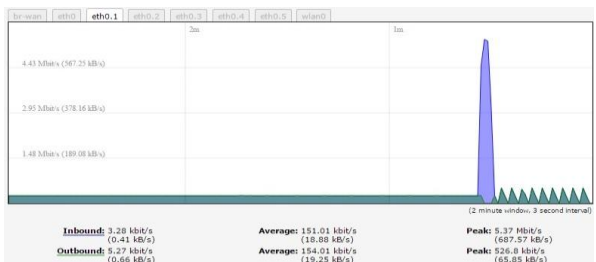

FIGURE 8: TRAFFIC MONITORING OF THE INTERFACE ETH2

That time presents in the Figure 8 is the necessary time to stop the attack and avoid compromising the OpenFlow network and validates the reaction mechanisms.

V. CONCLUSION

The Intrusion Detection and Treatment System (IDTS) proposal with its redirecting, mitigation and deceiving attacker security approach innovates and provides flexibilization, particularly in relation to the current intrusion detection using an IDS on a conventional style. This flexible technique, now associated with a network control mechanisms (SDN architecture/OpenFlow), allows an innovative way, not only to monitoring and detect potential attacks, as well as a "reaction" to the threats in a controlled and centrally manner. In effect, the validation tests reported illustrated only some of the possible protection approaches with the inclusion in the *whitelist* and reaction approaches with the inclusion in the *blacklist* and signaling to the administrator. Future works should include the use of others IDS on an architecture with more than one OpenFlow Switch to provide load balancing among other potential alternatives. The flexibility of the approach proposed reflects also the fact that we are proposing to secure a network with a logically centralized view and control although the controller itself might be fully distributed as commonly possible and frequently adopted on a SDN/OpenFlow network.